\documentclass[final,5p,times,twocolumn,authoryear]{elsarticle}

\usepackage{amsmath}
\usepackage{amssymb}
\usepackage{hyperref}
\usepackage{float}
\usepackage{multirow}
\usepackage{booktabs}

\hypersetup{
  colorlinks=true,
  linkcolor=blue,
  filecolor=magenta,
  urlcolor=cyan,
  pdftitle={Overleaf Example},
}
\usepackage{subfig}

\journal{\url{https://doi.org/10.1016/j.appet.2022.106096}}

\begin{document}

\begin{frontmatter}

  \title{Intake Monitoring in Free-Living Conditions: Overview and Lessons we
    Have Learned}

\author[hua]{Christos Diou}
\affiliation[hua]{organization={Department of Informatics and Telematics,  Harokopio University of Athens},
            country={Greece}}

\author[auth]{Konstantinos Kyritsis}
\author[auth]{Vasileios Papapanagiotou}
\author[auth]{Ioannis Sarafis}

\affiliation[auth]{organization={Electrical and Computer Engineering Department,
  Aristotle University of Thessaloniki},
            country={Greece}}



\begin{abstract}
  The progress in artificial intelligence and machine learning algorithms over
  the past decade has enabled the development of new methods for the objective
  measurement of eating, including both the measurement of eating episodes as
  well as the measurement of in-meal eating behavior. These allow the study of
  eating behavior outside the laboratory in free-living conditions, without the
  need for video recordings and laborious manual annotations. In this paper, we
  present a high-level overview of our recent work on intake monitoring using a
  smartwatch, as well as methods using an in-ear microphone. We also present
  evaluation results of these methods in challenging, real-world
  datasets. Furthermore, we discuss use-cases of such intake monitoring tools
  for advancing research in eating behavior, for improving dietary monitoring,
  as well as for developing evidence-based health policies. Our goal is to
  inform researchers and users of intake monitoring methods regarding (i) the
  development of new methods based on commercially available devices, (ii) what
  to expect in terms of effectiveness, and (iii) how these methods can be used
  in research as well as in practical applications.
\end{abstract}



\begin{keyword}
  Objective intake monitoring \sep Passive intake monitoring \sep Wearables
  \sep Artificial intelligence for dietary monitoring



\end{keyword}

\end{frontmatter}


\section{Introduction}
\label{sec:introduction}

Studying and monitoring food intake behavior is important for several health
conditions including obesity, eating disorders, cardiometabolic syndrome and
diabetes. Examples of methods for the in-depth study of intake include
video-based monitoring, where researchers manually annotate video sequences of
people eating (e.g., \cite{llewellyn2008eating}), and surface electromyography
(sEMG) where signals capture the activation of muscles associated with eating
(such as \cite{smit2011does}). These methods support detailed analysis of
intake behavior in the laboratory.

In free-living conditions, however, the options are limited to the use of
questionnaires, such as food recall questionnaires, food frequency
questionnaires or questionnaires reported via mobile apps. Self-reported food
intake information has been shown in many cases to be inaccurate, however,
while the use of self-reported information for policy decision-support purposes
has been questioned (see \cite{bingham1991limitations},
\cite{archer2013validity}). Furthermore, detailed in-meal eating parameters,
such as eating speed cannot be measured using self-reports.

As a result, the development of new sensors and algorithms for monitoring
intake behavior has received significant research attention over the past few
years. The main drivers for this interest have been the availability of
inexpensive sensors, the widespread use of mobile phones and wearable devices,
as well as recent developments in machine learning and signal processing
algorithms.

\begin{figure}
  \centering
  \includegraphics[width=.5\columnwidth]{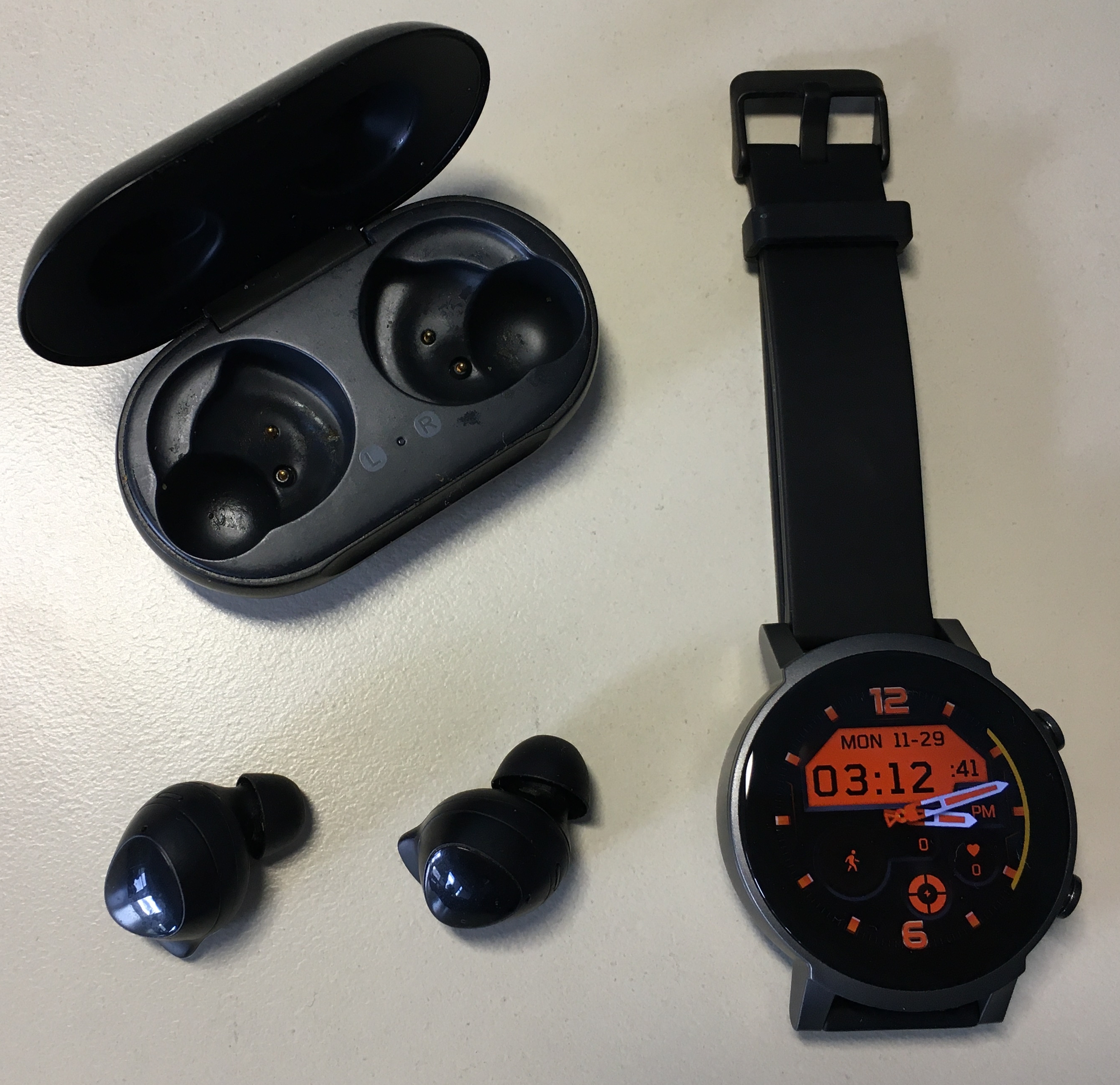}
  \caption{Modern commercially available hardware that can be used for
    objective eating behavior monitoring with the methods discussed in this
    paper: Bluetooth Earbuds and a Wear OS smartwatch.}
  \label{fig:wearables}
\end{figure}

In this paper, we present an overview of research that we have performed over
the past six years towards methods for passive, objective and accurate intake
monitoring in free-living conditions. Specifically, we focus on methods that use
sensors that are wearable and that are easy to acquire, possibly with
commercially available devices. We therefore present only methods for monitoring
intake using (a) ear-worn sensors (earbuds) and (b) wrist-worn sensors
(smartwatches). Figure \ref{fig:wearables} shows examples of commercially
available devices that can be used for intake monitoring using the methods
presented in this paper.

The goal of these methods is to directly obtain, or estimate, measurements that
can be broadly categorized in (a) in-meal and (b) all-day food intake
indicators. Examples are shown in Table \ref{tab:indicators}. All-day indicators
rely on accurate detection of meals and snacks. All other measurements can be
derived from this information. In-meal indicators can be more diverse and
require granular measurements. Furthermore, the proposed methods can be used to
study new features of intake behavior, possibly associated to health conditions,
as demonstrated by \cite{kyritsis2021assessment}.

\begin{table}
  \caption{Examples of in-meal and all-day intake indicators that can be
    measured using wearable sensors.}
  \label{tab:indicators}
  \centering
  \begin{tabular}{cc}
    \toprule
    \textbf{In-meal} & \textbf{All-day}\\ \midrule
    Meal duration & Main meals per day \\
    Bouts per minute & Number of snacks per day \\
    Bout deceleration & Eating breakfast \\
    Total intake & Eating schedule consistency\\ \bottomrule
  \end{tabular}
\end{table}

In Section \ref{sec:related_work} we present a brief overview of methods that
have been proposed in the recent literature for objective food intake monitoring
in free-living conditions, using various sensors. Section \ref{sec:inertial}
discusses several methods for monitoring intake using inertial sensors embedded
in a wrist-worn smartwatch. It is shown that it is not only possible to
explicitly model detailed movements performed during eating to measure intake,
but that it also is possible for deep representation learning algorithms to
learn the structure of eating movements implicitly.

Section \ref{sec:chewing} discusses methods for intake monitoring using chewing
detection. The methods were developed using a custom-made chewing detection
device that uses audio, photoplethysmography (PPG) and acceleration signals. It
is shown, however, that deep representation learning algorithms can achieve
similar results using the audio signal only. The algorithms therefore can be
used with commercially available earbuds that house a microphone.

The paper also discusses how these methods can be integrated into practical
use-cases for research, dietary monitoring and measurements for supporting
policy decisions (Section \ref{sec:usecases}). This includes a discussion on
the benefits of using passive objective monitoring, as well as potential
barriers and difficulties that are still under investigation.

\section{Related work}
\label{sec:related_work}

This section provides a brief overview of selected methods for
objective dietary monitoring that have been recently proposed in the
literature and are relevant to the work that we discuss in this paper.

\subsection{Bite detection}
\label{sec:bite_detection_literature}

\cite{dong2012new} presented a method that uses a single channel of
the gyroscope sensor (roll of the wrist) for detecting bites using a
smartwatch. This approach works well for counting the number of bites
but is not as accurate for temporal bite localization. Another
approach, proposed by \cite{zhang2016food}, uses a bottom-up solution,
aiming at first discovering subfeeding gestures (food-to-mouth and
back-to-rest) using the inertial signals and then aggregating these
gestures into bites. Both of these methods are further discussed in
Section \ref{sec:inertial}. Approaches also exist that bring the
cultural aspect into consideration when investigating eating
behavior. Such works make use of the characteristic wrist motions that
occur when certain cutlery is used. For example, the works of
\cite{cho2018asian} and \cite{kim2012analysis} use the accelerometer
data from wrist mounted sensors to deal with the recognition of
Asian-style eating events that mainly involve the use of the spoon and
chopsticks in contrast to the Western eating styles that mostly use
the fork and the knife.

The works of \cite{anderez2020eating} and \cite{amft2005detection}
showcase the feasibility of incorporating drinking gestures, in
addition to eating, to their models and perform in-meal monitoring
using wrist-mounted inertial sensors. Experimental results using small
datasets (involving six and two subjects, respectively) provide the
initial indication that simultaneously monitoring both gesture types
is possible; however, further experimentation is required using larger
and more diverse datasets in order to obtain more concrete evidence.

Other approaches include the use of video from a camera placed in
front of the participant during a meal session. Examples include
\cite{rouast2019learning, qiu2020counting} and
\cite{konstantinidis2020validation}. \cite{rouast2020single} proposed
a method that can be applied to both inertial and video data (but not
their combination) and is also able to distinguish between eating and
drinking episodes. Finally, \cite{heydarian2021exploring} presented a
method for the fusion of inertial and video data for in-meal eating
monitoring.

\subsection{Meal detection in-the-wild}

The approaches discussed in the previous section (section
\ref{sec:bite_detection_literature}) focus on measuring the in-meal
eating behavior by performing the temporal localization of bite events
via wrist motion tracking under free-living conditions. The current
section focuses on a much smaller body of research that aims at
identifying and localizing eating episodes (such as meals or snacks)
from data collected using wrist-mounted inertial sensors. Furthermore,
such approaches fall into two categories; the first category includes
methods that perform indirect meal detection by clustering
intake-related events (e.g., bites) while the second one includes
methods that perform direct detection.

Regarding the indirect detection of eating events, the work of
\cite{zhang2017generalized} presents a method that makes use of a
smartwatch for intake monitoring and a neck-worn camera for providing
ground truth. Their method is based on the rationale that most eating
activities do not occur while moving; thus, their proposed method
filters out periods with high physical activity and focuses on the
resulting ones. Using the density of the detected feeding gestures the
authors are able to detect the feeding episodes using a density-based
scheme. A similar method is presented in \cite{thomaz2015practical}
where the authors propose a pipeline that includes the extraction of
features from the inertial signals and a feeding gesture
classification scheme. Finally, eating moment detection is achieved by
using the DBSCAN algorithm on the detected food intake gestures.

Towards the direct detection of eating events, the work of
\cite{dong2011detecting} proposes the use of a wrist-band device and
the use of a single gyroscope channel by a two-state model (not eating
and possibly eating) based on thresholds. Their method is based on the
hypothesis that a period of increased wrist motion exists before and
after every meal, while during the meal the wrist motion energy is
decreased. The latter work of \cite{sharma2016automatic} extended the
work of \cite{dong2011detecting} by using a more well-suited novel
smartwatch-like sensing platform and increasing the dataset's size by
over twofold. Furthermore, the authors suggest that their initial
hypothesis (i.e., increased wrist activity before and after a meal)
may not generalize for all participants. Another work by
\cite{dong2013detecting} suggests the use of a prototype watch-like
device towards the detection of eating activities. More specifically,
the authors propose the use of a single channel from the on-board
gyroscope sensor, particularly the one that measures the orientation
velocity on the axis that is collinear with the forearm (also known as
the roll). By using the variance of the roll signal as the single
feature, they construct a finite state machine that consists of two
states, ``not eating'' and ``possibly eating'', while the transition
between those two states is achieved simply by
thresholding. \cite{mirtchouk2017recognizing} used two smartwatches,
one in each wrist, and recorded a large dataset (ACE), which is
publicly available. It includes meals, drinks and snacks. The work of
\cite{thomaz2017exploring} modeled eating as a bimanual task
(requiring two hands) and broke it down into two subcategories -
symmetric bimanual (hands have same role, e.g., holding a sandwich)
and asymmetric bimanual (hands have complementary roles, e.g., knife
and fork). Direct detection of eating events can also be achieved
using more complex sensors; e.g., \cite{jia2019automatic} as well as
\cite{raju2019processing} used a first-person, egocentric video stream
taken during the day to detect eating episodes.





\subsection{Chewing detection}

In addition to bite detection via eating gestures, several methods have explored
intake monitoring through chewing detection. \cite{4814940} proposed one of the
first intake monitoring systems that uses a combination of ear microphones (for
chewing) and throat microphones (for swallowing). \cite{amft_bite} suggests
bite-weight estimation from an in-ear microphone. The approach requires accurate
detection of chewing events, since the method is based on characteristics such
as chew duration, number of chews (per chewing bout), and signal energy during
the chew, to estimate bite weight. Unfortunately, these methods may not
generalize well in free-living conditions where external noise and movement
could interfere with detection.

\cite{passler7} also proposed methods which use an in-ear microphone, as well as
a second microphone for identifying external sounds. The authors propose several
algorithms that use mainly morphological and spectral features of the audio
signal. Again, generalization in noisy environments closer to real-life could be
an issue (see also Section \ref{sec:chewing}). In \cite{6556940}, the authors
compare eight different algorithms for chewing detection on 18 hours of
recordings. Best results achieve over $80\%$ both precision and recall. An
important finding is that the addition of a pre-processing stage where noise
(i.e. non-relevant sounds for chewing detection) is removed can reduce
false-positive detections by $28\%$.

Another method for chewing detection is presented by \cite{10.1145/3264902}, and
uses an ear-mounted sensor. The microphone is an off-the shelf device, however,
a 3D-printed mounting component is required, and authors report that it reduces
comfort during wear. Both temporal and spectral features are extracted from the
captured audio and a feature selection method is also used. The proposed method
detects $20$ - $24$ (depending on strictness of evaluation) eating episodes out
of a total of $26$ on a free-living dataset of $32$ hours.

\cite{7206521} uses a neck microphone. The prototype wearable is worn around the
neck, and a mounted microphone is mounted pressing against the throat, capturing
audio in an non-invasive way. Audio signal is transmitted via Bluetooth to a
smartphone were food-type recognition takes place. The proposed algorithm uses
Hidden Markov Model for chewing detection, and then extracts both time and
frequency domain features from the detected chewing segments. These features are
used with a decision-tree classifier to recognize food type. Classification
accuracy is reported to be $84.9\%$.

In \cite{6742586}, authors propose a combination of an in-ear microphone, a jaw
sensor, and wrist-mounted accelerometer for detecting eating activity. Different
features are extracted from each sensor and neural network classifiers are used
for eating detection. Authors report detection accuracy of $89.8\%$ on a $24$
hour dataset that includes both eating and daily activities. Alternative sensors
have also been examined; such a piezoelectric sensor is proposed in
\cite{6047558} for detecting chewing activity.

Recently, deep learning approaches have been used with audio sensors, as they
tend to achieve high effectiveness in most classification tasks.

One of the first methods to use deep learning for chewing signal processing is
the iHear Food system \cite{7545830}. The system relies on a
commercially-available headset for chewing detection. Authors evaluate SVM
classifiers for detection; they report accuracy of $94\%$ - $95\%$ in laboratory
conditions, which degrades to $65\%$ - $76\%$ in field testing. However, authors
then evaluate a deep learning classifier that can improve accuracy to $77\%$ -
$94\%$.

More recently, \cite{8551492} have suggested the use of a throat microphone,
combined with a convolutional neural network (CNN) after noise filtering. The
CNN is trained on spectral representation of the audio signal. Experimental
evaluation on $276$ minutes of recordings shows promising results.


\subsection{Food portion and food type recognition}

Complementary to methods for eating detection and in-meal eating behavior
monitoring, visual recognition methods have been proposed for assessing the
content and volume of food. For example, \cite{puri2009recognition} and
\cite{zhu2010use} propose the estimation of food portion size using
cameras. Further examples include \cite{kong2012dietcam, anthimopoulos2014food,
  zhu2011multilevel} and \cite{christodoulidis2015food}, that propose methods
for food type recognition from images. In \cite{s20154283} the authors proposed
the goFOOD$^{\text{TM}}$ system, which requires 2 photos or 1 short video of
food. It detects food, segments it and recognizes food type, and performs 3D
reconstruction for volume estimation. Then, it estimates calories and
micro-nutrients (based on a look-up table).

\section{Intake monitoring using smartwatches}
\label{sec:inertial}


This section presents how a smartwatch, used as a sensor platform, can
be utilized as a means of monitoring the in-meal and all-day eating
behavior. Albeit not technically a sensor, the smartwatch is a
wrist-mounted, watch-like device that contains sensors.
The eating behavior monitoring methods presented in this section make
use of the accelerometer and gyroscope sensors as they appear in the
majority of smartwatches due to their low cost and multiple
applications (e.g., adjusting the orientation of the screen).

Next, we introduce the \textit{eating behavior model} and its three
core components; in particular, the micromovement, the food intake
(FI) cycle (i.e., bite) and the meal. The term \textit{micromovement}
corresponds to a simple and short in duration movement of the wrist
that operates the utensil during the course of a meal. A typical
example of a micromovement is the upwards movement of wrist towards
the mouth area. In its ideal form, a \textit{food intake cycle} begins
by manipulating a utensil to pick up food from the plate (p),
continues with an upwards motion of the wrist operating the utensil
towards the mouth (u), followed by inserting the food into the mouth
(m) before concluding with a downwards motion of the wrist away from
the mouth area (d). In practice however, we observe repetitions of
certain micromovements, unrecognized wrist movements (o) or no wrist
movements at all (n).

\begin{figure}[h]
  \centering
  \includegraphics[width=0.50\textwidth]{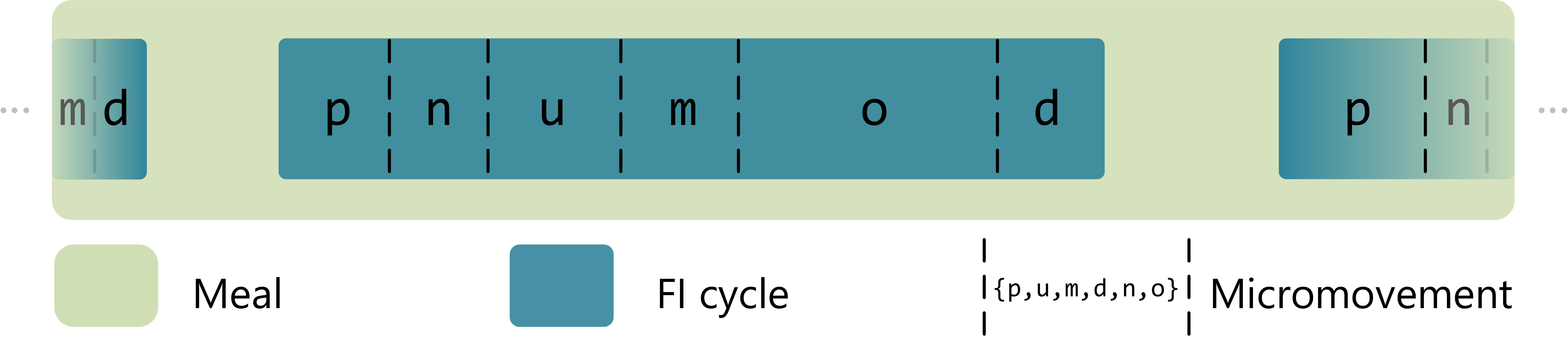}
  \caption{Example presenting a meal session (light green), a food
    intake cycle (cyan blue), and micromovements (between dotted
    vertical lines).}
  \label{fig:model}
\end{figure}

Finally, in the context of this work, the \textit{meal}
is defined as the sequence of actions that leads to the consumption of
already-prepared food, at a given place and lasts a finite amount of
time. Furthermore, we adopt the plated food assumption and that the
food is consumed using specific utensils (i.e., knife, fork and/or
spoon). Figure \ref{fig:model} depicts an example that makes use of
the proposed eating behavior model.

For the rest of this section, we present how the aforementioned eating
behavior model and the inertial data from a commercial smartwatch can
be used to perform in-meal, with and without the use of micromovements
(sections \ref{sec:micromov} and \ref{sec:microfree}), and all-day
eating behavior monitoring (section \ref{sec:allday}).


\subsection{Micromovement-based in-meal eating behavior monitoring}
\label{sec:micromov}

Given the micromovement-based eating behavior model introduced in
Section \ref{sec:inertial}, we propose a two-step approach towards the
detection of FI cycles during the course of a meal. The first step
deals with the processing of windowed inertial measurements (i.e.,
fixed-length segments) with the purpose of obtaining their
micromovement representations, while the second step is responsible
for modeling the temporal evolution and classify sequences of windows
as FI cycles.

The method's pipeline, originally introduced in \cite{kyritsis2019modeling},
begins with the preprocessing of the raw inertial data (3D acceleration and
angular velocity) using a median filter in order to smooth any short and
sudden fluctuations. In addition, since the accelerometer sensor captures both
the acceleration caused by the voluntary movement of the wrist as well as the
acceleration due to the Earth's gravitational field, we convolve the
accelerometer streams with a high-pass Finite Impulse Response (FIR)
filter. Following the preprocessing step, a Convolutional Neural Network (CNN)
is used to estimate the micromovement probability distribution of each window
of the inertial streams. To classify sequences of windows as FI cycles we apply
a Long short-term memory (LSTM) network. The LSTM network takes as input
overlapping sequences of micromovement distributions (as produced by the CNN)
and outputs a probability that the input sequence is an FI cycle. It should be
emphasized that the CNN and LSTM networks are trained \textit{separately}. More
specifically, the CNN is trained using windows of length equal to $0.2$ seconds
that correspond to the p, u, m, d, n micromovements, while the LSTM is trained
using sequences of length equal to $3.6$ seconds that contain micromovement
distributions that begin with p, end with d and contain an m micromovement
instance. The final FI cycle moments are assigned to the local maxima of the
LSTM network's output.

Earlier adaptations of the micromovement-based algorithm are based on
hand-crafted features and Support Vector Machines (SVM) instead of a
CNN \citep{kyritsis2017food} and Hidden Markov Models (HMM) for
modeling the temporal evolution of windowed micromovement instances
\citep{kyritsis2017automated}. These adaptations, however, have
proven to be subpar compared with the CNN and LSTM-based
approach. Finally, the adaptation presented in
\cite{papadopoulos2018personalised} includes the personalized tuning
of the SVM-based model \citep{kyritsis2017food} using unlabeled
samples and semi-supervised learning techniques.


\subsection{Micromovement-free in-meal eating behavior monitoring}
\label{sec:microfree}

Moving past the micromovement-based approach of Section
\ref{sec:micromov}, this section presents an end-to-end, single-step
method for the detection of FI cycles (i.e., bites) during the course
of a meal. It should be emphasized that the presented algorithm
depends solely on the information regarding the FI cycles and, as a
result, does not require the explicit knowledge of wrist
micromovements. In addition, this approach is considered as
data-driven since it processes the raw inertial data (not using
hand-engineered features) to obtain the micromovement representations
of the windowed inertial streams.

The core of the approach presented in \cite{kyritsis2020data} and
\cite{kyritsis2018end} is an Artificial Neural Network (ANN) with both
convolutional and recurrent layers. By allowing the model to be more data-driven the
CNN can discover the optimal feature representations before performing
temporal modeling using an LSTM.

\begin{figure*}[h]
  \centering
  \includegraphics[width=0.90\textwidth]{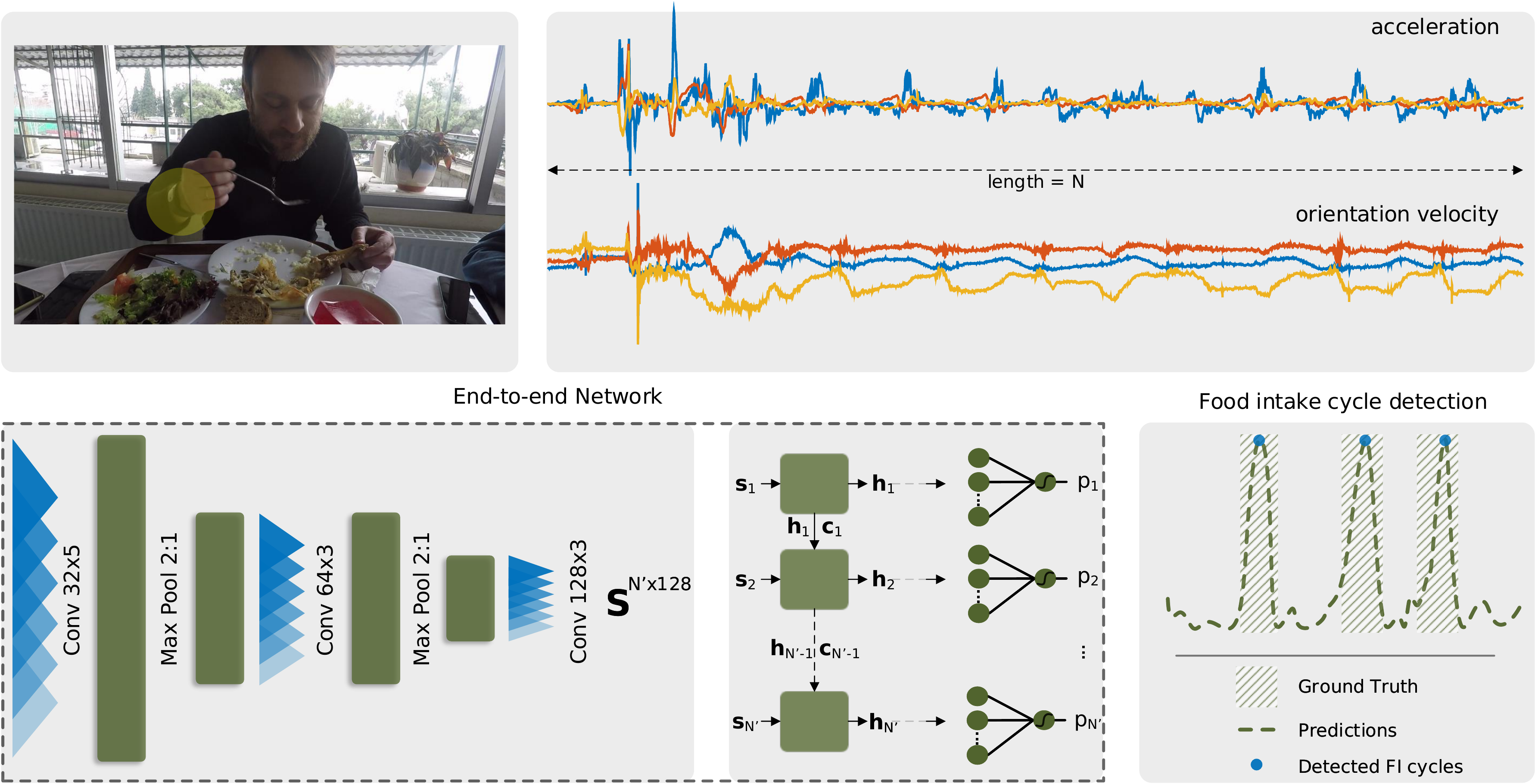}
  \caption{Overall pipeline of the end-to-end, bite detection
    approach. Initially, the preprocessed inertial signals that are captured
    during the meal are forwarded to the end-to-end network. The network
    processes the inertial data and outputs the bite probabilities. The variable
    $\mathbf{S}$ represents the intermediate output of the convolutional part of
    the network while $h_{i}$ and $c_{i}$ represent the output and cell state of
    the LSTM block.}
  \label{fig:e2e_arch}
\end{figure*}

Similar to Section \ref{sec:micromov}, preprocessing includes data smoothing and
removal of the gravitational component from the acceleration stream. The
end-to-end network is then trained by minimizing a \emph{single} loss function
(as opposed to the network of Section \ref{sec:micromov} that trained the CNN
and LSTM separately). Using signal samples of length equal to $5$ seconds, we
obtain the FI cycle predictions of the meal under investigation. To make the
model more robust, we propose a data augmentation technique that is based around
artificially changing the orientation of the smartwatch and thus simulating
different placements on the wrist. The final FI moments correspond to the local
maxima of the produced end-to-end network bite probabilities. Figure
\ref{fig:e2e_arch} depicts the architecture of the end-to-end network as part of
the in-meal FI cycle detection pipeline.

Evaluation is performed in a leave-one-subject-out (LOSO) fashion using our
strict evaluation scheme proposed by \cite{kyritsis2019modeling}. Essentially,
the first detected bite in a ground truth interval is considered as a true
positive (TP), while subsequent ones count as false positives (FP). Bite
detections outside ground truth intervals also count as FP. Ground truth
intervals without any bite detections are considered as false negatives
(FN). Furthermore, we used the publicly-available Food Intake Cycle (FIC)
dataset\footnote{\url{https://mug.ee.auth.gr/intake-cycle-detection/}}, containing a total of
$21$ meals from $12$ subjects recorded in the cafeteria of the Aristotle
University of Thessaloniki. The total duration of the meals sums to $246$
minutes, with a mean duration of $11.7$ minutes. No special instructions were
provided to the participants and they were free to consume a wide variety of
food types in their own way and pace and engage with discussions with other
people around the table or interact with their smartphone.

Table \ref{tab:table_results_fic_our} presents the evaluation
results. In addition to the micromovement-based or micromovement-free
approaches presented in Sections \ref{sec:micromov} and
\ref{sec:microfree}, we include experimental results from two
additional smartwatch-based methods found in the recent
literature. The method presented by \cite{dong2012new} is a
micromovement-free approach that is based on a single channel of the
gyroscope sensor (the one parallel to the forearm) and certain
thresholds to detect bite events. The method by \cite{zhang2016food}
is designed around the idea of modeling a bite using two subfeeding
wrist gestures; specifically, food-to-mouth and back-to-rest.

Evaluation results indicate that the micromovement-free, end-to-end
approach outperforms all other methods, both micromovement-based and
micromovement-agnostic, by achieving an F1 score of $0.928$ using our
strict evaluation scheme. The micromovement-based approach presented
in Section \ref{sec:micromov} achieves the second highest performance
with an F1 score of $0.913$ under the same scheme. Switching to the
less strict evaluation scheme of \cite{dong2012new} an overall
increase can be observed across all experiments. We emphasize on the
importance of using the F1 score as the evaluation metric of choice,
as it provides a combined measure of precision (how many of the
detected bites are actually bites) and recall (how many of the actual
bites are detected). More specifically, since the F1 score is the
harmonic mean of precision and recall, achieving a high F1 score
requires both a high precision \textit{and} a high recall.

\begin{table*}[ht]
  \centering
  \caption{FI cycle detection results in the form of TP, FP, FN,
    precision, recall and F1-score. The $\dagger$ symbol in the method
    proposed by \cite{dong2012new} is used to signify that parameter
    tuning was performed by optimization based on our strict
    evaluation scheme. The numbers inside the parentheses represent
    the results obtained using the relaxed evaluation scheme proposed
    by \cite{dong2012new}. The MM column indicates whether the method is
    based in wrist micromovements or not.}
  \label{tab:table_results_fic_our}
  \scalebox{0.90}{
    \begin{tabular}{l| c c c c c c c}
      \toprule
      \textbf{Method} & \textbf{MM (y/n)} & \textbf{TP} & \textbf{FP} & \textbf{FN} &  \textbf{Prec} & \textbf{Rec} & \textbf{F1} \\ \midrule
      \cite{kyritsis2020data} & n  & $1,231$ ($1,237$) & $102$ ($96$) & $101$ ($95$) & $0.923$ ($0.927$) & $0.924$ ($0.928$) & $\mathbf{0.923}$ ($0.928$) \\
      \cite{kyritsis2017food}& y  & $1,221.5$ ($1,267.6$) & $228.4$ ($182.3$)& $110.5$ ($64.4$) & $0.842$ ($0.874$)& $0.917$ ($0.951$)& $0.878$ ($0.911$)\\
      \cite{kyritsis2019modeling}& y & $1,241.8$ ($1,263.4$) & $144.5$ ($122.9$) & $90.2$ ($68.6$) & $0.895$ ($0.911$) & $0.932$ ($0.948$)& $0.913$ ($\mathbf{0.929}$)\\
      \cite{zhang2016food}& y  & $944$ ($1,102$) & $431$ ($233$) & $388$ ($230$) & $0.686$ ($0.825$) & $0.708$ ($0.827$)& $0.697$ ($0.826$) \\
      \cite{dong2012new}& n & $707$ ($1,190$)& $794$ ($311$) & $625$ ($142$)& $0.471$ ($0.792$)& $0.530$ ($0.893$)& $0.499$ ($0.840$)\\
      \cite{dong2012new} $\dagger$ & n & $772$ ($1,214$) & $746$ ($304$)& $560$ ($118$)& $0.508$ ($0.799$) & $0.579$ ($0.911$)& $0.541$ ($0.851$) \\ \bottomrule
    \end{tabular}
  }
  \end{table*}

\subsection{All-day and in-the-wild eating behavior monitoring}
\label{sec:allday}

A more challenging problem arises when dealing with all-day,
in-the-wild recordings where meals are only a small part of the
recorded data, while the rest of the recording corresponds to
everyday activities such as commuting, doing sports or working. This
section addresses the problem of ``when'' one eats by performing the
temporal localization of \textit{meals} in all-day, in-the-wild
recordings.

The hypothesis around the design of our method is that the density of
bites is high during meals and low when outside of meals. Based on
this hypothesis, the first step of our approach
(\cite{kyritsis2020data, kyritsis2019detecting}) is to detect
\textit{all} FI cycles that occur during during the all-day
recording. The next step is to use signal processing techniques to
isolate regions with high bite density and, iteratively, merge them
with adjacent ones (within $180$ seconds of each other). Furthermore,
any regions with duration less than $180$ seconds are discarded (short
meal rejection). The outcomes of the above, two-step, process are the
recording's meal end-points.


Evaluation is performed by exhaustively calculating the TP, FP, FN and
true negatives (TN) by discretizing the complete all-day
recording. Moreover, we are able to measure the intersection and union
between the detected and ground truth meal intervals, thus allowing to
calculate the Jaccard index (also known in the literature as
Intersection over Union - IoU). The datasets used for evaluation are the
publicly available FreeFIC (both the baseline and held-out parts are
available
online\footnote{\url{https://mug.ee.auth.gr/free-food-intake-cycle-detection/}})
and ACE (proposed by \cite{mirtchouk2017recognizing}). The baseline
part of FreeFIC contains a total of $16$ all-day recordings adding up
to $77.32$ hours, collected from $6$ subjects. Similarly, the held-out
part of FreeFIC consists of $6$ all-day recordings, from $6$ subjects,
with a total duration of $35.39$ hours. On the other hand, ACE
contains $25$ recordings from $11$ subjects summing up to
approximately $250$ hours. One significant difference between the two
datasets is that ACE includes the consumption of drinks and snacks
which have different wrist mechanics compared to the utensil-based
consumption of meals that are included in FreeFIC.

The first half of Table \ref{tab:meal_results} summarizes the
performance of meal detection algorithms obtained in a LOSO fashion
using the baseline FreeFIC set, while the second half using the
held-out part of FreeFIC. We compare the performance of our approach
with the well-known method of \cite{dong2013detecting} - a
segmentation algorithm based on the hypothesis that meals tend to be
preceded and succeeded by periods of vigorous wrist motion. Also,
inspired by \cite{zhang2016food}, we compare the performance of the
DBSCAN clustering algorithm towards the aggregation of FI cycles into
meals. Results from both LOSO and held-out experiments point out that
the proposed approach outperforms similar methods by achieving a
weighted accuracy/Jaccard index equal to $0.953$/$0.820$ and
$0.964$/$0.821$, respectively. By applying our meal detection
algorithm to the ACE dataset (Table \ref{tab:meal_results_ace}), we
can see that despite the different recording conditions the weighted
accuracy remains satisfactory ($0.788$ and $0.825$ with and without
considering snacks as meals, respectively). Since eating activities
occupy a very small portion of the total recording duration, therefore
it is a problem that by definition involves imbalanced datasets, it is
crucial to adjust accuracy by a weight factor to avoid bias in the
results. For the meal detection problem, the weight factor is obtained
by dividing the total recording duration with the time spent during
meals; e.g., for the FreeFIC dataset the weight factor is calculated
as $77.32/5.42=14.2$, which essentially means that recordings contain
$14.2$ times more non meal-related activities that meal-related ones.

\begin{table*}[ht]
  \centering
  \caption{Meal detection results using the FreeFIC dataset, performed
    in a LOSO and held-out fashion.}
  \label{tab:meal_results}
  \scalebox{0.90}{
  \begin{tabular}{l| l c c c c c c c }
    \toprule
    \textbf{Experiment} & \textbf{Method} &  \textbf{Prec} & \textbf{Rec} & \textbf{Spec} & \textbf{F1} & \textbf{Acc} & \textbf{Acc}$_{w}$ & \textbf{JI} \\ \midrule
    \multirow{3}{*}{LOSO} & \cite{kyritsis2020data} &$0.880$ & $0.919$&$0.990$ &$0.899$ &$0.985$ & $\mathbf{0.953}$&$0.820$ \\
    & {DBSCAN} & $0.838$ & $0.895$ & $0.986$ & $0.865$ & $0.980$ & $0.939$ & $0.752$ \\
    & \cite{dong2013detecting} & $0.323$  &$0.525$  & $0.919$ &$0.400$ &$0.892$  &$0.545$ &$0.255$ \\ \midrule

    \multirow{3}{*}{Held-out}& \cite{kyritsis2020data} &$0.858$ & $0.937$&$0.992$ &$0.896$ &$0.990$ & $\mathbf{0.964}$&$0.821$ \\
    & {DBSCAN} & $0.774$ & $0.779$ & $0.989$ & $0.776$ & $0.979$ & $0.882$ & $0.681$ \\
    & \cite{dong2013detecting} & $0.105$  &$0.697$  & $0.714$ &$0.182$ &$0.717$  &$0.182$ &$0.089$ \\ \bottomrule
  \end{tabular}
  }
\end{table*}

\begin{table*}[ht]
  \centering
  \caption{Meal detection results obtained by the
    \cite{kyritsis2020data} algorithm using the ACE dataset.}
  \label{tab:meal_results_ace}
  \scalebox{0.90}{
  \begin{tabular}{l c c c c c c c}
    \toprule
    \textbf{Positive periods} & \textbf{Prec} & \textbf{Rec} & \textbf{Spec} & \textbf{F1} & \textbf{Acc} & \textbf{Acc}$_{w}$ & \textbf{JI} \\ \midrule
    Meals only  & $0.397$ & $0.710$ & $0.934$  & $0.509$ & $0.921$ & $0.825$ & $0.346$ \\
    Meals \& snacks & $0.457$ & $0.633$ & $0.939$ & $0.531$ & $0.917$ & $0.788$ & $0.377$ \\ \bottomrule
  \end{tabular}
  }
  \end{table*}

\subsection{Limitations}
\label{sec:discussion_inertial}

The relatively high battery requirements from recording accelerometer and
gyroscope signals can impose limitations on intake monitoring with smartwatches
that are currently available commercially. In an attempt to overcome these
issues, we are investigating the use of the accelerometer signal only for the
detection of FI cycles and localization of meals. Early experiments indicate
that, despite the significant loss of information, the bite and meal detection
results remain satisfactory, yet open to improvement. Specifically, LOSO
evaluation using FIC reached an F1 score of $0.713$ for bite detection and a
weighted accuracy/Jaccard index of $0.884$/$0.636$ using FreeFIC for meal
detection.

Future work includes the investigation of how different sitting postures and
use of utensils affect micromovement modeling, bite detection and extracted
indicator. In addition, we plan on incorporating drinking gestures to the
eating behavior model, to make it capable of monitoring more complex eating
behaviors.

\section{Chewing detection and measurement}
\label{sec:chewing}

A different type of signal that can be used to automatically extract eating
behavior information is audio. In particular, continuously capturing audio using
a microphone close to the mouth can be used to capture the distinct sounds of
crushing food between the teeth during mastication and thus detect chewing
activity.

Using audio and detecting chewing offers much richer information regarding
eating behavior. In particular, it is possible to identify individual chews and
thus extract information, such as number of chews per chewing bout, chewing rate,
changes in chewing rate, and more. Furthermore, it is possible to extract
additional information, such as food texture attributes (e.g. crispiness and
moisture) which could help in identifying the food type. Finally, approaches to
estimate the weight (in grams) of each chewing bout are also possible (and are
currently under investigation).

The most common placing of microphones for such applications in inside
the outer in-ear canal. Early studies (such as \cite{10.1007/11551201_4}) have
shown that this placement is one of the most effective ones as chewing sounds
travel very well (with very low attenuation) from the mandible, through the
skull, to the ear. In addition, as the microphone is usually directed towards
the inside of the ear canal, it naturally dampens external sounds (such as
environmental noise, other people talking, and more). On the other hand, sounds
such as talking (from the person wearing the microphone sensor) or walking
introduce significant challenges for the detection systems.

\begin{figure}
  \centering
  \includegraphics[width=.7\columnwidth]{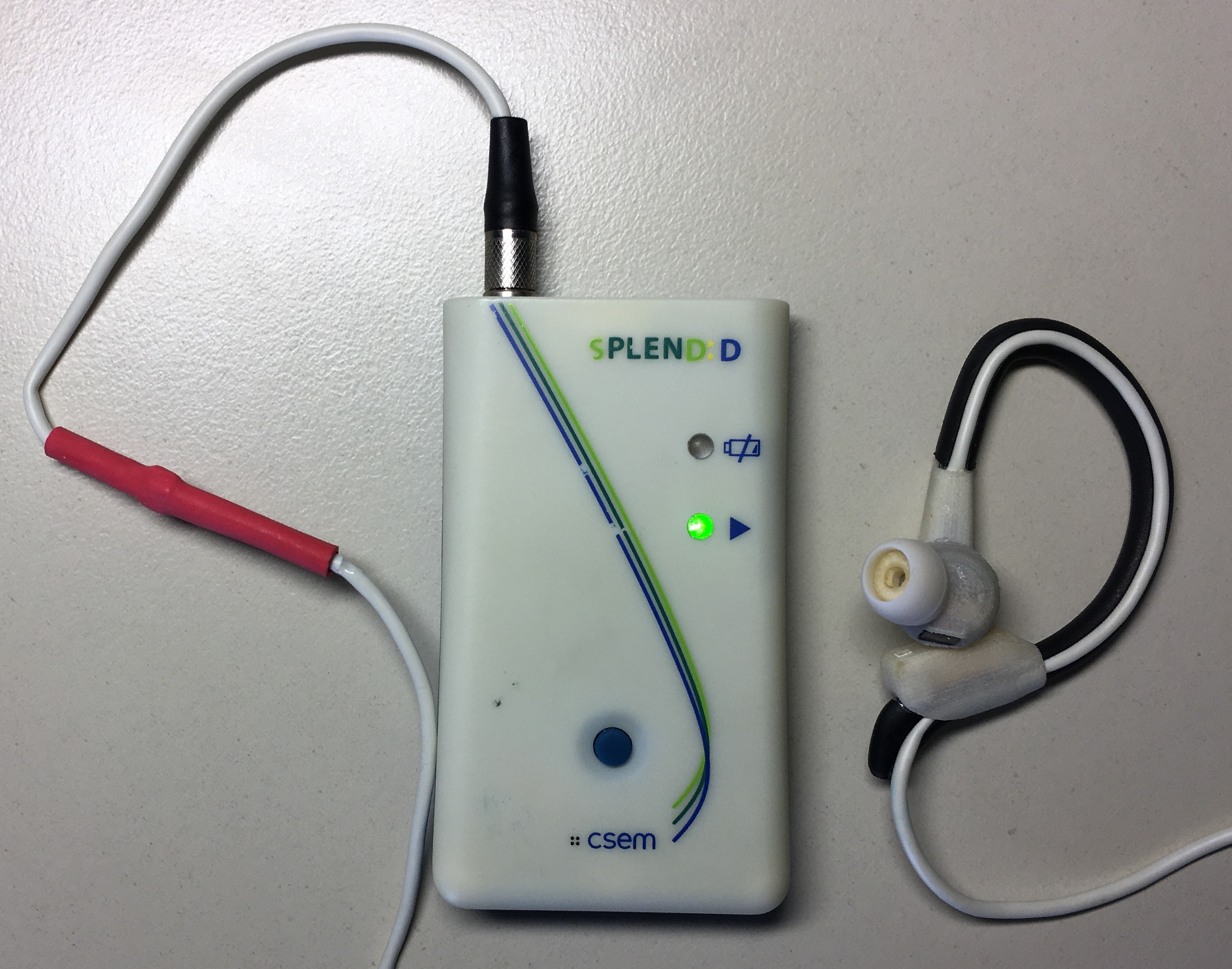}
  \caption{The prototype chewing detection sensor, developed in the context of
    the EU SPLENDID project.}
  \label{fig:splendid_sensor}
\end{figure}

A complete approach for a ``chewing sensor'' has been developed in the context
of the EU funded SPLENDID project \citep{maramis2014preventing}. In this
project, a prototype sensor (hardware and software) has been developed that
enables objective and automatic monitoring of chewing activity. The sensor
includes two pieces: an ear-mounted device, that connects via wire to a
belt-mounted device (Figure \ref{fig:splendid_sensor}). The ear-mounted device
includes an in-ear microphone and a photoplethysmography (PPG) sensor. The
belt-mounted device includes a triaxial accelerometer, a data logger that
buffers signals from all three sensors and a Bluetooth transmitter that
transfers chunks of the buffered data to an Android smartphone. The design of
the sensor improved over the course of the project, focusing on the
comfort of the users, since the ear-piece needs to be worn continuously. It has
received mostly positive comments during the project's pilot evaluations
\cite{chewing-feasibility}.

While microphones had been studied and experimented with in the literature, the
use of a PPG for chewing detection had not been reported before. PPG sensors
usually detect blood-flow related information. For chewing detection, placement
of a PPG sensor around the ear lobe can detect fluctuations in blood flow that
are caused by the activation of the masseter and the trigeminal nerve during
chewing \citep{ppg}.

\subsection{Chewing detection using audio, PPG, and acceleration signals}

The chewing detection algorithm by \cite{jbhi-combo} follows a late-fusion
approach where each of the audio and PPG signals are processed independently,
and their final detection scores are aggregated for the final decision.

Both signals are first pre-processed by applying a low-pass filter to remove
unwanted drift that is caused by the electronics and hardware. Then, the signals
are split into sliding windows. Different window lengths are chosen for each
signal, based on the information we want to obtain. Specifically, we use windows
of $5$ s length for the PPG signal, since we want to extract the rhythmic motion
of chewing that typically is in the range from $1$ to $3$ Hz. Thus, even for the
lowest frequency of $1$ Hz, a window of $5$ s contains at least $5$ periods,
i.e. $5$ individual chews. From each window we extract spectral features that
capture the energy of frequencies relevant to chewing frequency. A trained SVM
model is then used on the feature vectors to obtain a classification score.

A similar process is followed for the audio signal. Here, however, we extract
much shorter windows ($100$ and $200$ ms) and extract features that capture the
\textit{texture} of the sound, i.e. the various characteristics of sound as food
is crushed between the teeth. We extract spectral features, higher-order
statistics, and other features such as fractal dimension that help the
classification model distinguish chewing from talking \citep{fractal}. A trained
SVM model is used to obtain a classification score for each window. Since
audio-based windows are extracted at a higher rate compared to PPG-based
windows, a down-sampling process is required to synchronize the two
classification scores.

The scores, $s_{\text{PPG}}$ and $s_{\text{audio}}$ are combined using a single
tuning parameter and chewing is decided using the following inequality
\begin{equation}
  \label{eq:chewing_late_fusion}
  s_{\text{PPG}}[n] + \alpha \cdot s_{\text{audio}}[n] > A_{\text{fusion}}
\end{equation}
where parameter $\alpha$ defines the relative importance of audio over the PPG signal
on the decision, and parameter $A_{\text{fusion}}$ is a threshold that defines
the strictness of classification.

The reason for using parameter $\alpha$ is that audio and PPG have different
responsiveness in the detection process. Specifically, PPG has very high recall
at the cost of many false-positive detections. On the other hand, audio-based
detection yields very good precision (very few false-positives) but has lower
recall. Thus, the combination of the two parameters allows the user to tune the
detection system towards higher recall or precision, or a balanced mode, based
on the needs of each use-case.

The accelerometer signal is complementary used to detect periods of walking,
running, and high physical activity. This information is important because
walking rhythm is similar to chewing (i.e. $1$ to $3$ Hz) and it registers
similar patterns both to the audio and PPG sensors, causing the classification
models to yield high false-positive detections. By filtering out the chewing
detections during walking or other high physical activities we can remove these
false positives. It can be argued that one can eat and walk at the same time,
but overall, the gain in chewing detection effectiveness is greater when the
accelerometer signal is used.

The system and algorithm has been validated on various data collection studies
during the SPLENDID project. The final version has been evaluated on a large and
challenging dataset that was collected in the Wageningen University in the
Netherlands and includes approximately $60$ h of recordings. Participants
consumed a variate of food types during both meals and snacks, and were able to
freely move and leave the premises, and were also encouraged to engage in
activities with high physical-activity level. The dataset has been made publicly
available \citep{splendid-dataset}, including all signals and manual annotations
for each individual chew. The only exception is the audio signals which have not
been published due to privacy issues, but we have made available all the
audio-based features we have extracted.

\begin{table}
  \centering
  \caption{Chewing detection results using audio, PPG, and acceleration signals
     \citep{jbhi-combo}. Results are obtained for $\alpha=1$ (equal
    importance between audio and PPG) and for classification threshold
    $A_{\text{fusion}}$ a little over $0.5$.}
  \label{tab:chewing-detection-results-svm}
  \begin{tabular}{lccccccc}
    \toprule
    \textbf{} & \textbf{Prec} & \textbf{Rec} & \textbf{Acc} & \textbf{Acc}$_w$ & \textbf{F1} \\
    \midrule
    LOSO       & $0.79$ & $0.81$ & $0.94$ & $0.89$ & $0.51$ \\
    Cumulative & $0.70$ & $0.80$ & $0.93$ & $0.88$ & $0.75$ \\
    \bottomrule
  \end{tabular}
\end{table}

\subsection{Audio-feature learning for chewing detection}

Selecting which features to extract from a window is based on careful analysis
of the signals, their nature, and the expertise of the researchers. An
alternative method is to use methods that automatically learn to extract
relevant features during the training of the classification model.

We have examined this possibility by employing deep neural networks (DNNs) to
detect chewing based only on the audio signal of the sensor
\citep{chewing-cnn}. We have studied the effectiveness of a DNN that includes
(a) five convolutional layers that are used to learn meaningful and effective
representations and (b) three fully connected layers that discriminate between
chewing and non-chewing based on the learned features of the CNN layers. We have
experimented with various design choices for the network architecture, and
\cite{chewing-cnn} presented the results (on the same dataset of
\cite{splendid-dataset}) for five different architectures with different lengths
of input window (from $1$ to $5$ s). The most effective one is the $5$ s input
window architecture, achieving an F1-score of $0.908$ which is higher than the
comparable $0.748$ of the SVM-based approach (Table
\ref{tab:chewing-detection-results-svm}).

\begin{table}
  \centering
  \caption{Chewing detection results using only audio and the $5$ s input window
    length DNN, (a) using fully-supervised learning \citep{chewing-cnn} and (b)
    using self-supervised learning (and a non-linear projection head during only
    the unsupervised training) \citep{chewing-ssl}.}
  \label{tab:chewing-detection-results-cnn}
  \begin{tabular}{cccccccc}
    \toprule
    & \textbf{Prec} & \textbf{Rec} & \textbf{Acc} & \textbf{Acc}$_w$ & \textbf{F1} \\
    \midrule
    (a) & $0.89$ & $0.93$ & $0.98$ & $0.96$ & $0.91$ \\
    (b) & $0.94$ & $0.79$ & $0.97$ & $0.92$ & $0.86$ \\
    \bottomrule
  \end{tabular}
\end{table}

While the DNN-based models achieve higher effectiveness, they require
significantly larger volumes of labeled data. Labeling data is a laborious and
slow process, that can also sometimes introduce errors. To alleviate this
process, we have explored the possibility of using unlabeled data to train the
convolutional layers of the model in an unsupervised way, and then use labeled
data for supervised training of the fully connected layers only
\citep{chewing-ssl}. This self-supervised learning approach has been found to be
highly promising, and our first results indicate that self-supervised learning
can lead to comparable, and sometimes even better, detection effectiveness
(Table \ref{tab:chewing-detection-results-cnn}). These results demonstrate that
it is possible to rely on the audio signal only and to use commercially
available earbuds instead of purpose-made chewing sensors, as in our previous
work by \cite{bite-weight}.

\subsection{Using audio to extract in-meal information}

Having a system that continuously monitors and detects chewing activity can be
used to extract additional information besides the total duration of chewing
(which is what we evaluate in Tables \ref{tab:chewing-detection-results-svm} and
\ref{tab:chewing-detection-results-cnn}). Individually detected chews are
aggregated into chewing bouts, and these in turn into meals/snacks. As a result,
we can extract information such as total number of meals/snacks, duration of
each meal, etc. It is also possible to extract more detailed information within
each meal, such as number of chewing bouts (within a meal), number of individual
chews (per chewing bout or for the entire meal), chewing rate, and more. The
accuracy of this information is directly tied to the effectiveness of the chewing
detection models.

Furthermore, in \cite{food-texture-attributes} we have presented a
modification of our algorithm by \cite{jbhi-combo}, which, given chewing
detections, can estimate certain food-texture attributes. These include
crispiness, wetness/moisture, and chewiness. A preliminary evaluation on a
relatively small dataset has shown very promising results, especially for
crispiness.

We have also presented an algorithm that estimates the weight (in grams) of each
food bite based only on the audio signal \citep{bite-weight}. The method is
similar to the one by \cite{jbhi-combo} and extracts the same features, however,
it then uses different regression models to estimate bite-weight. Both
food-type--specific and non-specific models have been evaluated, and the best
results (using regression neural networks) achieve an average absolute
error of less than $1$ gram per bout, which corresponds to less than $10\%$
relative error for $3$ out of the $4$ food types examined.





\section{Use-cases and impact in eating behavior research and practice}
\label{sec:usecases}

The intake monitoring solutions that we have seen so far can be used in both
experimental studies,
as well as in real-world applications at individual or population level. In
this section, we outline three categories of use-cases, and discuss practical
considerations resulting from previous experience in the SPLENDID
\citep{maramis2014preventing}, i-Prognosis \citep{hadjidimitriou2016active} and
BigO \citep{diou2020bigo} projects.

\subsection{Use in research}

The study of intake behavior often requires the use of specialized measuring
equipment or manual annotation through video recordings and is therefore
commonly performed inside the laboratory. Examples include the study of in-meal
eating behavior (e.g. \citep{ioakimidis2009method, zandian2009linear}) and the
study of mastication behavior (e.g., \citep{woda2006adaptation}). In
free-living conditions, studies of intake behavior commonly rely on
self-reports that are either provided once (such as food recall or food
frequency questionnaires) or through multiple ``just-in-time'' user feedback
through mobile apps \cite{maramis2019}.

For detecting intake moments, measuring eating via a smartwatch or through the
chewing sensor using the methods presented in Sections \ref{sec:inertial} and
\ref{sec:chewing} can increase recall in eating events. This is valuable for
snacking and constant eating behaviors, that take place unconsciously
\citep{wansink2007mindless} and are not accurately reported. Furthermore, one
can have reliable estimates of the error introduced by the measurement methods,
as reported in Tables \ref{tab:meal_results} and
\ref{tab:chewing-detection-results-cnn}.

For in-meal behavior monitoring, the smartwatch-based method of Section
\ref{sec:inertial} can provide the number of bites and the bite frequency, as
well as other eating parameters which cannot be easily self-reported. An example
demonstrating the value of gesture-based measurements for research was shown by
 \cite{kyritsis2021assessment}. In this study, the micromovement approach was
used to quantify eating difficulties of Parkinson's Disease patients. The study
analyzed in-meal eating behavior of Parkinson's Disease patients as well as
healthy subjects, both inside and outside the clinic, for approximately 700
meals. It was found that the duration of the movement from the plate to the
mouth during eating was a reliable, objective measurement that was able to
distinguish Parkinson's Disease patients from healthy subjects.

\paragraph*{Impact summary} The presented intake monitoring methods enable (a)
 automated measurements of in-meal indicators without videos and without manual
 effort by researchers, (b) objective and more accurate intake behavior
 measurements, compared to self-reports and (c) automated intake behavior
 measurements in free-living conditions.

\subsection{Use for patient dietary monitoring}

In addition to research, the proposed objective intake monitoring tools have
the potential to be used for measuring the eating behavior of patients in
free-living conditions. This enables targeted recommendations and
improved treatment for a number of health domains, such as overweight and
obesity, diabetes, cardiometabolic syndrome and cardiovascular disease.

\begin{figure}
  \includegraphics[width=\linewidth]{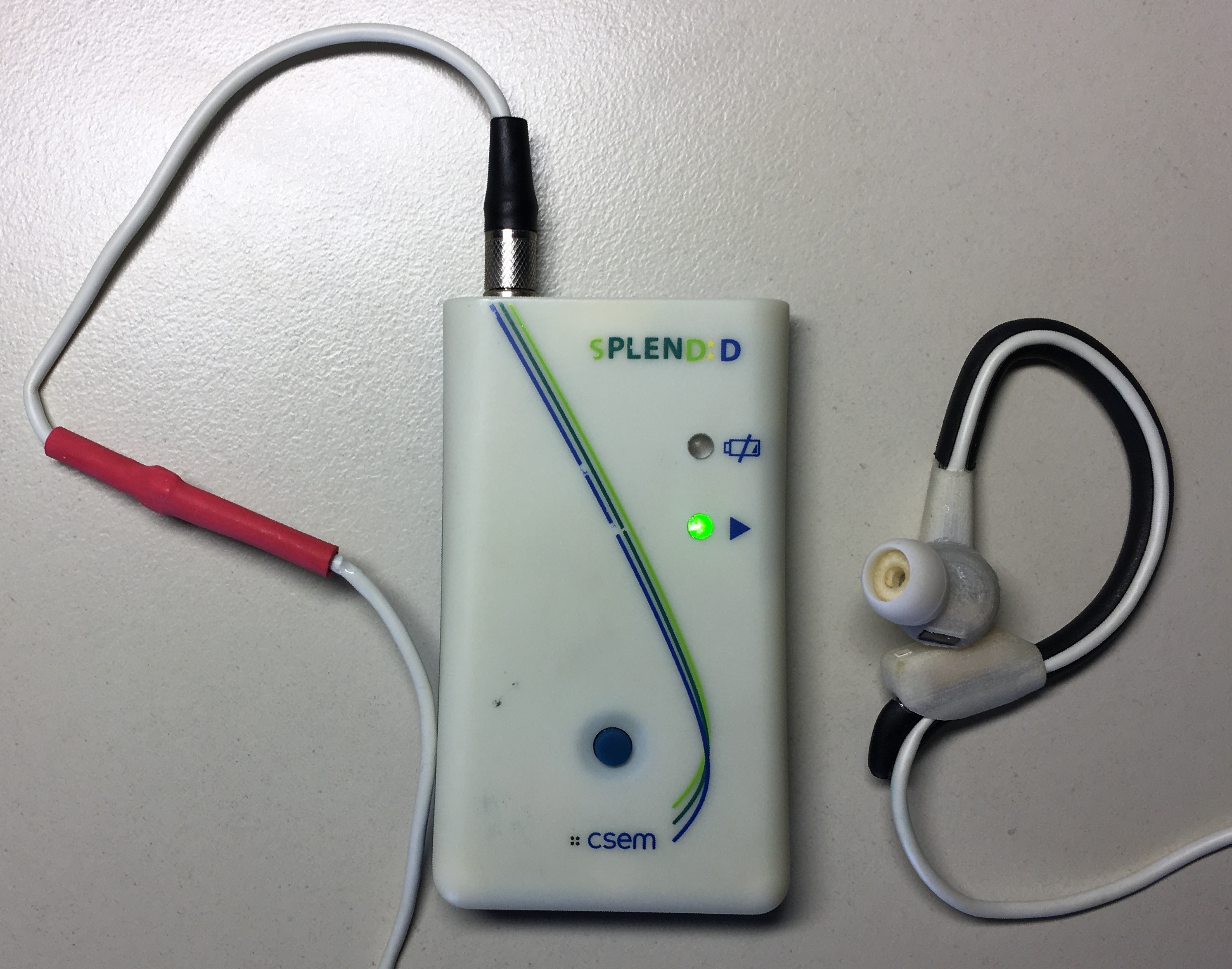}
  \caption{A view of the SPLENDID dietary monitoring system. Eating occurrences
    and physical activity of a subject over a number of selected days}
  \label{fig:splendid_dashboard}
\end{figure}

In this scenario, patients are instructed to use the monitoring tools under
prespecified conditions (e.g., for at least 3 hours when at home in the
afternoon), or even freely throughout the day. The devices collect signals which
are then analyzed to extract behavioral indicators, such as the ones shown in
Table \ref{tab:indicators}. These are presented to the health professional who
is monitoring the patient through a web application. The indicators can be
presented in the form of aggregated summaries (e.g., average number of afternoon
snacks) or through detailed timeseries graphs, bar plots and other
visualizations. Such an example visualization, from the SPLENDID dietary
monitoring system, is shown in Figure \ref{fig:splendid_dashboard}.

In the SPLENDID project this approach was used in a series of studies with the
chewing sensor (\cite{chewing-feasibility}) to (a) support sensor development,
(b) assess its usefulness and acceptance by end-users and health professionals
and (c) evaluate the feasibility of use in real-world settings. In a study
where 20 overweight young adults used the chewing sensor integrated prototype
for 4 weeks, it was found that the sensor was useful, but there are important
usability limitations, especially when it comes to long-term use. For example
users reported that the cable of the prototype was inconvenient, that the sensor
was conspicuous (and thus users avoided its use in a social context), while its
use for several hours was uncomfortable. Today, wireless earbuds housing an
integrated microphone, such as the ones shown in Figure \ref{fig:wearables} are
commercially available, and therefore several of these drawbacks have already
been addressed.

\paragraph*{Impact summary} Use of the proposed sensors by patients in
free-living conditions allows improved, objective monitoring by clinicians, who
can identify unhealthy behaviors (such as excessive snacking, constant eating,
increased eating speed) and provide targeted recommendations.

\subsection{Use for policy decision support}

The proposed algorithms can work with ``off-the-shelf'' devices sold
commercially, such as smartwatches and wireless earbuds. This enables their
large-scale use, at population level, to support health policy decisions. As a
motivating example, imagine a local health authority that wants to address the
increasing obesity prevalence in the young population, through interventions in
the local urban environment and through regulation of fast food restaurants in
the city (e.g., by limiting the number of permits close to schools). Ideally,
these decisions should be supported by evidence on the effectiveness of these
interventions, (a) to ensure that resources are well allocated, and (b) to
provide quantifiable arguments in support of the interventions.

To achieve this, the health authorities invite citizens to participate in a
one-month study (possibly providing incentives) by installing a mobile app. The
mobile app should not send any raw, personally identifiable data to the
authorities (e.g., by extracting indicators at the individual's mobile phone),
while at the same time providing valuable information about the association of
individual's behavior with the local environment.

Use of intake monitoring combined with location in this scenario can provide
answers to questions related to the eating behavior of individuals with respect
to their environment. For example, it could answer questions, such as ``where
do children have their main meals? In the school, at restaurants, or at
home?'', or ``How much do they eat when visiting restaurants and cafeterias
close to school?'', ``Do they eat multiple snacks at or around school?'', ``Is
their behavior significantly different when they eating at home versus eating
outside?'' Note that to answer such questions, one does not need to send the
exact locations visited by the individuals. It is possible to only send behavior
indicator statistics about the different location types (e.g., ``school'',
``home'', ``cafe'', ``restaurant''), without specifying where these locations
are.

A similar approach was followed in the BigO project (\cite{filos2021exploring,
  diou2020bigo}), in which approximately 4000 children and adolescent participants used
their mobile phones to measure physical activity behavior, as well as
statistics related to visits to various location types. The intake information
provided by the users of BigO was only based on pictures and self-reports, but
would have been much richer if the proposed intake measurement methods had also
been applied.

\paragraph*{Impact summary} The proposed intake behavior
measurement methods can be applied at population level with equipment that is
currently commercially available at low cost. It is therefore possible to
collect detailed evidence about determinants of the urban and socioeconomic
environment on eating habits, as well as on the effectiveness of policies and
interventions to improve population health.

\subsection{Current barriers and future research directions}


Objective, passive and continuous food intake monitoring methods show
significant promise in improving the amount and quality of data produced for the
purposes of research, patient guidance and policy decision support. On the other
hand, with the current state of the art there are still significant obstacles
that need to be overcome in order to facilitate further deployment of these
technologies in practice.

Among the barriers, the most important ones include user acceptance, compliance
and deployment with existing hardware and operating systems. Acceptance is both
related to user comfort, as well as what is socially acceptable. For example,
the experimental integrated prototype of the chewing sensor was not
well-received outside the home environment, since it was not fully incospicuous
and could raise questions among young peers. The development of more acceptable
devices such as smartwatches that look like regular watches, or wireless
earbuds with integrated microphones, will hopefully improve user acceptance.

Acceptance is a prerequisite for compliance, but is not sufficient. Compliance
depends on additional factors, such as the effect of the algorithms on battery
duration as well as the users remembering to wear the devices when
instructed. Differences in compliance introduce challenges in the analysis of
the data, since only partial measurements are available for each user during
the day. Drawing reliable inferences about the behavior of individuals in the
presence of such types of ``missing data'' is one important current research
problem.

Finally, in the cases where ``off-the-shelf'' equipment is used to support the
use-cases outlined above, the restrictions imposed by the vendors' operating
system may introduce significant difficulties in the implementation and
deployment of intake monitoring applications. For example, the operating system
may not provide access to raw accelerometer or gyroscope signals, or it may
forcefully stop applications that perform processing of the signals. These are
technical restrictions that result from the fact that commercially available
devices (mobile phones, smartwatches) have not been designed for continuous
data collection. We believe that as the benefits of such data acquisition
functionalities become more apparent, it will become easier to implement and
deploy monitoring applications for real-world use cases.


\section{Conclusions}
\label{sec:conclusions}

In this paper we have presented a high-level overview of methods for objective
food intake monitoring with commercially available - and usable - sensors. We
have provided an overview of recent work performed by our group, along with
results demonstrating the effectiveness compared to baseline methods. We have
also discussed various use-cases, benefits and limitations of such intake
monitoring systems. Our overall aim has been to inform readers regarding what
is possible when off-the-shelf sensors are combined with algorithms based on
modern machine learning methods.

It was shown that by using an off-the-shelf smartwatch it is possible to acquire
highly accurate detection of intake moments during a meal, reaching an F1 score
of over $0.92$. This approach can be extended to detect meals, reaching an F1
score of $0.9$ and an accuracy of $0.99$ (weighted accuracy of $0.96$) in the
FreeFIC dataset. Meal detection results are worse in the ACE dataset ($0.79$
weighted accuracy when including snacks, $0.82$ for main meals), however they
can still be considered satisfactory. Potential drawbacks of the smartwatch
methods include the need for wearing the smartwatch on the wrist that is used
to transfer food to the mouth, as well as the fact that the algorithm needs to
constantly record the accelerometer as well as the gyroscope signal. Finally,
most meals in the training and evaluation datasets have been performed using a
fork or spoon and a knife.

Regarding chewing detection, we showed that using a combination of signals
(audio, PPG and accelerometer) can lead to accurate detection of chews using a
combination of frequency and time domain features and a decision fusion
mechanism. Results on a challenging real-world dataset with over 60 hours of
recordings showed that the method achieved a weighted accuracy of $0.89$. Using
the audio signal only, after representation learning with a deep convolutional
neural network achieved even better results, reaching a weighted accuracy of
$0.96$. Importantly, it was shown that this approach also works with recent
self-supervised learning methods, which can learn useful representations of the
chewing signals without the use of labeled examples for model training. These
results are important, not only with respect to their effectiveness, but
because in-ear microphones are now available in several commercial wireless
earbuds, allowing these methods to be deployed using inexpensive, off-the-shelf
devices. The largest challenge for this family of methods remains usability and
compliance, since users will inevitably wear their earbuds only for a part of
the day. Methods for estimating intake behavior when only a subset of the data
is available is still a subject of ongoing research.

During the last years, we have also used these methods for a number of
applications, including research on intake behavior for patients with
Parkinson's Disease, dietary monitoring of individuals with overweight or
obesity, as well as population monitoring for developing targeted policies
against obesity. These use-cases revealed that there are still many
difficulties to overcome when using these methods in operational environments,
including technical/equipment difficulties, usability, user compliance as well
as the analysis and interpretation of results. On the other hand, these
challenges highlight new opportunities for future research, that we believe
will bring intake monitoring methods to the everyday practice.

\section*{Acknowledgements}
The work leading to these results has received funding from the European Commission
under projects SPLENDID (Grant Agreement No. 610746, 01/10/2013-30/09/2016),
BigO (Grant Agreement No. 27688, 01/12/2016 - 30/11/2020), iPrognosis (Grant
Agreement No. 690494, 01/02/2016 - 31/01/2020) and REBECCA (Grant Agreement
No. 965231, 01/04/2021 - 31/03/2025).

\bibliographystyle{elsarticle-harv} 
\bibliography{biblio}

\end{document}